\documentclass[11pt]{article}
\usepackage{epsfig}
\def\nn{\nonumber}
\def\be{\begin{equation}}
\def\ee{\end{equation}}
\newcommand{\bea}{\begin{eqnarray}}
\newcommand{\eea}{\end{eqnarray}}
\newcommand{\bdm}{\begin{displaymath}}
\newcommand{\edm}{\end{displaymath}}
\long\def\symbolfootnote[#1]#2{\begingroup%
\def\thefootnote{\fnsymbol{footnote}}\footnote[#1]{#2}\endgroup}

\def\sdbeta{s_{2\beta}}

\def\sq2{\sqrt{2}}
\def\drbar{\overline{\rm DR}}
\def\msbar{\overline{\rm MS}}
\def\smalldrbar{\scriptscriptstyle{\overline{\rm DR}}}

\def\gl{\tilde{g}}
\def\mg{m_{\gl}}
\def\g{\mg^2}
\def\gq{\mg^4}

\def\del{\Delta}
\def\hsm{H_{\scriptscriptstyle{\rm SM}}}
\def\gam{Z}

\newcommand{\smallz}{{\scriptscriptstyle Z}} 
\newcommand{\smallw}{{\scriptscriptstyle W}} %
\newcommand{\smallr}{{\scriptscriptstyle R}} %
\newcommand{\mz}{m_\smallz}

\newcommand{\muF}{\mu_{\scriptscriptstyle F}}
\newcommand{\muR}{\mu_\smallr}

\def\mt{m_t}
\def\stu{\tilde{t}_1}
\def\std{\tilde{t}_2}

\def\t{\mt^2}
\def\tq{\mt^4}
\def\tu{m_{\tilde{t}_1}^2}
\def\td{m_{\tilde{t}_2}^2}
\def\ti{m_{\tilde{t}_i}^2}
\def\tuq{m_{\tilde{t}_1}^4}
\def\tdq{m_{\tilde{t}_2}^4}
\def\sdt{s_{2\theta_t}}
\def\cdt{c_{2\theta_t}}

\def\mb{m_b}
\def\sbu{\tilde{b}_1}
\def\sbd{\tilde{b}_2}

\def\bi{m_{\tilde{b}_i}^2}
\def\bl{m_{\tilde{b}_{\scriptscriptstyle L}}^2}
\def\br{m_{\tilde{b}_{\scriptscriptstyle R}}^2}

\def\biq{m_{\tilde{b}_i}^4}
\def\sdb{s_{2\theta_b}}

\newenvironment{appendletterA}
 {
  \setcounter{section}{0}
  \setcounter{equation}{0}
  
 }{
 }

\begin{document}

\begin{titlepage}


{\flushright{
        \begin{minipage}{5cm}
	  CERN-PH-TH/2008-124\\
          RM3-TH/08-12 \\
	  LAPTH-1254/08
        \end{minipage}        }

}
\renewcommand{\thefootnote}{\fnsymbol{footnote}}
\vskip 2cm
\begin{center}
\boldmath
{\LARGE\bf On the NLO QCD corrections \\[7pt]
to Higgs production and decay in the MSSM}\unboldmath
\vskip 1.cm
{\Large{G.~Degrassi$^{a,\,b}$ and P.~Slavich$^{a,\,c}$}}
\vspace*{8mm} \\
{\sl ${}^a$ CERN, Theory Division, CH-1211 Geneva 23, Switzerland}
\vspace*{2.5mm}\\
{\sl ${}^b$
    Dipartimento di Fisica, Universit\`a di Roma Tre and  INFN, Sezione di
    Roma Tre \\
    Via della Vasca Navale~84, I-00146 Rome, Italy}
\vspace*{2.5mm}\\
{\sl ${}^c$  LAPTH, 9, Chemin de Bellevue, F-74941 Annecy-le-Vieux,  France}
\end{center}
\symbolfootnote[0]{{\tt e-mail:}}
\symbolfootnote[0]{{\tt degrassi@fis.uniroma3.it}}
\symbolfootnote[0]{{\tt Pietro.Slavich@cern.ch}}

\vskip 0.7cm

\begin{abstract}
We present explicit analytic results for the two-loop top/stop/gluino
contributions to the cross section for the production of CP-even Higgs
bosons via gluon fusion in the MSSM, under the approximation of
neglecting the Higgs boson mass with respect to the masses of the
particles circulating in the loops. The results are obtained employing
the low-energy theorem for Higgs interactions adapted to the case of
particle mixing. We discuss the validity of the approximation used by
computing the first-order correction in an expansion in powers of the
Higgs boson mass. We find that, for the lightest CP-even Higgs boson,
the gluino contribution is very well approximated by the result
obtained in the limit of vanishing Higgs mass. As a byproduct of our
calculation, we provide results for the two-loop QCD contributions to
the photonic Higgs decay.
\end{abstract}
\vfill
\end{titlepage}    
\setcounter{footnote}{0}


\section{Introduction}

One of the main goals of the present experimental program at the
Tevatron and at the Large Hadron Collider (LHC) is the search for the
Higgs boson(s) in order to elucidate the mechanism of  electroweak
symmetry breaking. To support this goal, an accurate theoretical
knowledge of the Higgs production cross-section, its decay modes, and
the important background processes is required (for a general review
see ref.~\cite{Dj}).

In the Standard Model (SM) the minimal realization of the Higgs sector
predicts a single neutral Higgs boson $\hsm$, whose mass can be
constrained by electroweak precision data and the direct-search limit
form LEP to be lighter than roughly 200 GeV.  At LHC the main
production mechanism for $\hsm$ is the loop-induced gluon fusion
mechanism \cite{H2gQCD0}, $gg \to \hsm$, where the coupling of the
gluons to the Higgs is mediated by loops of colored fermions,
primarily the top quark. The knowledge of this process in the SM
includes the full next-to-leading order (NLO) QCD corrections
\cite{H2gQCD1,SDGZ}, the next-to-next-to-leading order (NNLO) QCD
corrections in the limit of infinite top mass \cite{H2gQCD2},
soft-gluon resummation effects \cite{bd4}, an estimate of the
next-to-next-to-next-to-leading order (NNNLO) QCD effects
\cite{Moch:2005ky} and also the first-order electroweak corrections
\cite{DjG,ABDV0}.

The Minimal Supersymmetric extension of the Standard Model, or MSSM,
features a richer Higgs spectrum that consists of two neutral CP-even
bosons $h, H$, one neutral CP-odd boson $A$ and two charged scalars
$H^\pm$. As in the SM, the gluon-fusion process is the main production
mechanism for the neutral Higgs bosons.  Furthermore, in the MSSM, the
coupling of the gluons to the Higgs bosons is mediated not only by
colored fermions but also by their supersymmetric partners. Thus, the
study of the NLO QCD corrections to the cross section for Higgs boson
production via gluon fusion in the MSSM requires the investigation of
a larger variety of diagrams with respect to the SM case. Two-loop
diagrams involving squarks and gluons were first considered in
ref.~\cite{Dawson:1996xz} under the assumption that the Higgs boson
mass can be neglected w.r.t.~the masses of the particles running in
the loops. Later this approximation has been relaxed and the full
dependence on all the relevant masses has been retained
\cite{babis1,ABDV,MS}.  Explicit two-loop results for the diagrams
that involve quarks, squarks and gluinos, as well as for the diagrams
that involve the D-term-induced quartic squark couplings, have been
presented so far only in the limit of vanishing Higgs mass. Indeed,
ref.~\cite{HS1} provided analytic formulae for the contributions
involving top quark and/or stop squarks, valid for vanishing Higgs
mass and under the additional simplifying limits of zero stop mixing
and hierarchical patterns of soft SUSY-breaking masses.
Ref.~\cite{HS2} presented a computation of the same contributions
valid for arbitrary parameters in the stop sector. However, the
explicit results were too long to be printed, and were made available
only in the form of a computer code. Recently, the rather daunting
calculation of the full two-loop QCD Higgs-gluon-gluon amplitude in
the MSSM for arbitrary Higgs mass has been completed \cite{babis2}.
The calculation relies on a combination of analytical and numerical
methods, and explicit results have not been made available so far.

Given the importance of the Higgs-physics search program, it is highly
desirable to have the NLO radiative corrections to the gluon-fusion
cross section expressed in a simple and flexible analytic form that
can be easily used to investigate large regions of the MSSM parameter
space.  To this purpose, we present explicit analytic results for the
two-loop top/stop/gluino contributions to the cross section for
CP-even Higgs boson production, under the approximation of neglecting
the Higgs mass with respect to the masses of the particles circulating
into the loops. Our results were obtained by adapting the low-energy
theorem for Higgs interactions \cite{LET1} to the case of particle
mixing. We confirm the result of ref.~\cite{HS2} by a numerical
comparison, after accounting for the different renormalization scheme
adopted in that paper. We then discuss the validity of the
approximation of vanishing Higgs boson mass by computing the first
correction to it, i.e.~terms of ${\cal O}(m_h^2/M^2)$ where $M$ is the
generic mass of the particles circulating in the loop. We show that,
for the lightest Higgs boson $h$, the gluino contribution is very well
approximated by the result obtained in the limit of vanishing Higgs
mass. As a byproduct of our calculation, we provide results for the
two-loop QCD contributions to the photonic Higgs decay.

The paper is organized as follows: in section \ref{sec:general} we
summarize general results on the cross section for Higgs boson
production via gluon fusion. In section \ref{sec:results} we describe
our calculation of the two-loop contributions to the CP-even
Higgs-gluon-gluon form factors.  In section \ref{sec:corr} we assess
the validity of the approximation of vanishing Higgs mass. In section
\ref{sec:decays} we explain how to adapt our results to the
calculation of the gluonic and photonic decay widths of the Higgs
bosons. In section \ref{sec:concl} we discuss the applicability to
other processes of our way to compute the SUSY-QCD
corrections. Finally, in the appendix we provide the explicit
analytical results for the two-loop top/stop/gluino contributions to
the Higgs-gluon-gluon form factors.



\section{Higgs production via gluon fusion at NLO in the MSSM }
\label{sec:general}

In this section we summarize some general results on Higgs boson
production via gluon fusion. The hadronic cross section for Higgs
boson production at center-of-mass energy $\sqrt{s}$ can be written as
\be
\sigma(h_1 + h_2 \to \phi+X)  \,=\, 
          \sum_{a,b}\int_0^1 dx_1 dx_2 \,\,f_{a,h_1}(x_1,\muF)\,
         f_{b,h_2}(x_2,\muF)  \times
\int_0^1 dz~ \delta \left(z-\frac{\tau_\phi}{x_1 x_2} \right)
\hat\sigma_{ab}(z)~,
\label{sigmafull}
\ee
where $\phi=(h,H)$, $\tau_\phi= m^2_\phi/s$, $\muF$ is the
factorization scale, $f_{a,h_i}(x,\muF)$, the parton density of the
colliding hadron $h_i$ for the parton of type $a, \,(a = g,q,\bar{q})$
and $\hat\sigma_{ab}$ the cross section for the partonic subprocess $
ab \to \phi +X$ at the center-of-mass energy $\hat{s}=x_1 \,x_2\,
s=m^2_\phi/z$. The latter can be written in terms of the leading-order
(LO) contribution $\sigma^{(0)}$ as
\be
\hat\sigma_{ab}(z)=
\sigma^{(0)}\,z \, G_{ab}(z) \, .
\label{Geq}
\ee
 
We consider now the production of the lightest CP-even Higgs boson,
$h$, through gluon fusion. The LO term can be written as
\be
\sigma^{(0)}  =  
\frac{G_\mu \,\alpha_s^2 (\muR)  }{128\, \sqrt{2} \, \pi}\,
\left|T_F \left( -\sin\alpha \,{\mathcal H}^{1\ell}_1 +
\cos\alpha \,{\mathcal H}^{1\ell}_2 \right)  \right|^2~,
\label{ggh}
\ee
where $G_\mu$ is the muon decay constant, $\alpha_s(\muR)$ is the
strong gauge coupling expressed in the $\overline{\rm MS}$
renormalization scheme at the scale $\muR$, $T_F=1/2$ is a color
factor, and $\alpha$ is the mixing angle in the CP-even Higgs sector
of the MSSM. ${\mathcal H}_i$ ($i = 1,2$) are the form factors for the
coupling of the neutral, CP-even component of the Higgs doublet $H_i$
with two gluons, which we decompose in one- and two-loop parts as
\be
{\mathcal H}_i ~=~ {\mathcal H}_i^{1\ell}
~+~ \frac{\alpha_s}{\pi} \,  {\mathcal H}_i^{2\ell}
~+~{\cal O}(\alpha_s^2)~.
\label{Hdec}
\ee
The coefficient function $G_{ab}(z)$ in eq.~(\ref{Geq}) can in turn be
decomposed, up to NLO terms, as
\be
G_{a b}(z)  ~=~  G_{a b}^{(0)}(z) 
~+~ \frac{\alpha_s}{\pi} \, G_{a b}^{(1)}(z) ~+~{\cal O}(\alpha_s^2)\, ,
\label{Gdec}
\ee
with the LO contribution given only by the gluon-fusion channel:
\bea
G_{a b}^{(0)}(z) & = & \delta(1-z) \,\delta_{ag}\, \delta_{bg} \, .
\eea
The NLO terms include, besides the $gg$ channel, also the one-loop
induced processes $gq \rightarrow qh$ and $q \bar{q} \rightarrow g h$:
\bea
G_{g g}^{(1)}(z) & = & \delta(1-z) \left[C_A \, \frac{~\pi^2}3 
\,+ \,\beta_0 \, \ln \left( \frac{\muR^2}{\muF^2} \right) 
 \,+ \,\left(\frac{-\sin\alpha \,{\mathcal H}^{2\ell}_1 
+\cos\alpha \,{\mathcal H}^{2\ell}_2}{
-\sin\alpha \,{\mathcal H}^{1\ell}_1 
+ \cos\alpha \,{\mathcal H}^{1\ell}_2 } 
\, +  \,{\rm h.c.} \right)  \right]  \nn \\
&+ &  P_{gg} (z)\,\ln \left( \frac{\hat{s}}{\muF^2}\right) +
    C_A\, \frac4z \,(1-z+z^2)^2 \,{\cal D}_1(z) +  C_A\, {\cal R}_{gg}  \, , 
\label{real}
\eea

\be
G_{q \bar{q}}^{(1)}(z) ~=~   {\cal R}_{q \bar{q}} \, , ~~~~~~~~~~~
G_{q g}^{(1)}(z) ~=~  P_{gq}(z) \left[ \ln(1-z) + 
 \frac12 \ln \left( \frac{\hat{s}}{\muF^2}\right) \right] + {\cal R}_{qg} \,,
\label{qqqg}
\ee
where the LO Altarelli-Parisi splitting functions are
\be
P_{gg} (z) ~=~2\,  C_A\,\left[ {\cal D}_0(z) +\frac1z -2 + z(1-z) \right]
\label{Pgg} \, ,~~~~~~~~~~~
P_{gq} (z) ~=~  C_F \,\frac{1 + (1-z)^2}z~. 
\ee
In the equations above, $C_A =N_c$ and $C_F = (N_c^2-1)/(2\,N_c)$
($N_c$ being the number of colors), $\beta_0 = (11\,
C_A - 2\, N_f)/6 $ ($N_f$ being the number of active flavors) is the
one-loop $\beta$-function of the strong coupling in the SM, and
\be
{\cal D}_i (z) =  \left[ \frac{\ln^i (1-z)}{1-z} \right]_+  \label {Dfun} \, .
\ee

The $gg$-channel contribution, eq.~(\ref{real}), involves two-loop
virtual corrections to $g g \rightarrow h$ and one-loop real
corrections from $ gg \to h g$. The former, regularized by the
infrared singular part of the real emission cross section, are
displayed in the first row of eq.~(\ref{real}).  The second row
contains the non-singular contribution from the real gluon emission in
the gluon fusion process.  The function ${\cal R}_{gg}$ is obtained
from one-loop diagrams where only quarks or squarks circulate into the
loop, and in the limit in which the Higgs boson is much lighter than
the particles in the loop it goes to ${\cal R}_{gg} \to -11 (1-z)^3/(6
z)$.
Similarly, the functions ${\cal R}_{q \bar q}$ and ${\cal R}_{q g}$ in
eq.~(\ref{qqqg}) describe the $ q \bar q \to h g $ annihilation
channel and the quark-gluon scattering channel, respectively. They are
obtained from one-loop quark and squark diagrams, and in the
light-Higgs limit they go to ${\cal R}_{q \bar q} \to 32 \,(1-z)^3/(27
z)$,~ ${\cal R}_{q g} \to 2\,z/3 - (1-z)^2/z$.
The functions ${\cal R}_{gg},\, {\cal R}_{q \bar q},\,
{\cal R}_{q g}$ are actually completely known. Their expressions  can be 
obtained from the results of ref.~\cite{BDV} (see also refs.~\cite{ehsv,BG}).

The one-loop form factors ${\cal H}_1^{1\ell}$ and ${\cal
H}_2^{1\ell}$ contain contributions from diagrams involving quarks or
squarks. The two-loop form factors ${\cal H}_1^{2\ell}$ and ${\cal
H}_2^{2\ell}$ contain contributions from diagrams involving quarks,
squarks, gluons and gluinos.  Focusing on the contributions involving
the third-generation quarks and squarks, and exploiting the structure
of the Higgs-quark-quark and Higgs-squark-squark couplings, we can
write to all orders in the strong interactions
\bea
{\mathcal H}_1 & = & \lambda_t \,\left[
\mt \,\mu\,\sdt\,F_t
\,+ \mz^2 \,\sdbeta \,D_t  \right] \;+ 
\lambda_b \,\left[\mb\,A_b\,\sdb\,F_b \,+  2\,\mb^2\,G_b \,+ 
2\, \mz^2 \,c_\beta^2 \,D_b
\right]\,, \label{eq:H1} \\
{\mathcal H}_2 & = & \lambda_b\,\left[
\mb \,\mu\,\sdb\,F_b
\,-\mz^2 \,\sdbeta \,D_b  \right] + 
\lambda_t\, \left[
\mt\,A_t\,\sdt\,F_t \,+  2\,\mt^2\,G_t \,- 
2\, \mz^2 \,s_\beta^2 \,D_t
\right]\label{eq:H2}~.
\eea
In the equations above $\lambda_t = 1/\sin\beta$ and $\lambda_b =
1/\cos\beta$, where $\tan\beta \equiv v_2/v_1$ is defined as the ratio
of the vacuum expectation values (vev) of the neutral components of
the two Higgs doublets. Also, $\mu$ is the higgsino mass parameter in
the MSSM superpotential, $A_q$ (for $q=t,b$) are the soft
SUSY-breaking Higgs-squark-squark couplings and $\theta_q$ are the
left-right squark mixing angles (here and thereafter we use the
notation $s_\varphi \equiv \sin\varphi, \, c_\varphi \equiv
\cos\varphi$ for a generic angle $\varphi$). The functions $F_q$ and
$G_q$ appearing in eqs.~(\ref{eq:H1}) and (\ref{eq:H2}) denote the
contributions controlled by the third-generation Yukawa couplings,
while $D_q$ denotes the contribution controlled by the electroweak,
D-term-induced Higgs-squark-squark couplings. The latter can be
decomposed as
\be
D_q  = \frac{I_{3q}}2 \, \widetilde{G}_q
+ c_{2\theta_{\tilde q}} \, 
\left(\frac{I_{3q}}2 -  Q_q \,s^2_{\theta_\smallw} 
\right) \,\widetilde{F}_q \, , \label{eq:Dq}
\ee
where $I_{3q}$ denotes the third component of the electroweak isospin
of the quark $q$, $Q_q$ is the electric charge and $\theta_\smallw$ is
the Weinberg angle.

The one-loop functions entering ${\mathcal H}_i^{1\ell}$ are
\bea
F_q^{1\ell} ~=~ \widetilde F_q^{1\ell}
& =&  \frac{1}{2}\,\left[
\frac1{m^2_{\tilde{q}_{1}}} {\mathcal G}^{1\ell}_{0}(\tau_{\tilde{q}_{1}}) -
\frac1{m^2_{\tilde{q}_{2}}} {\mathcal G}^{1\ell}_{0}(\tau_{\tilde{q}_{2}}) 
\right]\, , \label{eq:F1l}\\
G_q^{1\ell} & =& \frac{1}{2}\,\left[
\frac1{m^2_{\tilde{q}_{1}}} {\mathcal G}^{1\ell}_{0}(\tau_{\tilde{q}_{1}}) +
\frac1{m^2_{\tilde{q}_{2}}} {\mathcal G}^{1\ell}_{0} (\tau_{\tilde{q}_{2}}) + 
\frac1{m_q^2} {\mathcal G}^{1\ell}_{1/2} (\tau_q)\right]~, \label{eq:G1l}\\
\widetilde G_q^{1\ell} & =& \frac{1}{2}\,\left[
\frac1{m^2_{\tilde{q}_{1}}} {\mathcal G}^{1\ell}_{0} (\tau_{\tilde{q}_{1}}) +
\frac1{m^2_{\tilde{q}_{2}}} {\mathcal G}^{1\ell}_{0} (\tau_{\tilde{q}_{2}}) 
\label{eq:Gtilde}\right]~,
\eea
where $\tau_k \equiv 4\,m_k^2/m_h^2$, and the functions ${\mathcal
G}^{1\ell}_{0}$ and ${\mathcal G}^{1\ell}_{1/2}$ read
\bea
{\mathcal G}^{1\ell}_{0} (\tau) & =& ~~~~\,\tau \!\left[ 1 + \frac{\tau}{4}\, 
 \log^2 \left(\frac{\sqrt{1- \tau} - 1}{\sqrt{1- \tau} + 1}\right) \right]\,,
\label{eq:4} \\
{\mathcal G}^{1\ell}_{1/2} (\tau) & = & - 2\,\tau
 \left[ 1 - \frac{ 1 -\tau}4  \,   
 \log^2 
  \left(\frac{\sqrt{1-\tau} - 1}{\sqrt{1-\tau} + 1} \right) \right] \,.
\label{eq:3}
\eea
The analytic continuations are obtained with the replacement $m_h^2
\rightarrow m_h^2 + i \epsilon$~. For later convenience, we remark
that in the limit in which the Higgs boson mass is much smaller than
the mass of the particle running in the loop, i.e.~$\tau\gg 1$, the
functions ${\mathcal G}^{1\ell}_{0}$ and ${\mathcal G}^{1\ell}_{1/2}$
behave as
\be
{\mathcal G}^{1\ell}_{0} \rightarrow -\frac13 -\frac{8}{45\,\tau} 
~+~{\cal O}(\tau^{-2})~,~~~~~~~~~~
{\mathcal G}^{1\ell}_{1/2} \rightarrow -\frac43-\frac{14}{45\,\tau} 
~+~{\cal O}(\tau^{-2})~.
\label{Glimit}
\ee

The discussion above has focused on the production of the lightest
CP-even Higgs mass-eigenstate $h$, whose mass is bounded at tree-level
by the $Z$-boson mass and can hardly exceed 130--140 GeV when
radiative corrections are taken into account (see, e.g.,
ref.~\cite{higgsbound}). In this mass range, it can be expected that
the exact values of the functions ${\cal H}_1^{2\ell}$ and ${\cal
H}_2^{2\ell}$ are well approximated by the results computed in the
limit in which the $h$ boson mass is neglected w.r.t.~the masses of
the particles running in the loops, which we are going to call the
vanishing Higgs-mass limit (VHML). In the next section we derive
explicit analytic results for the two-loop top/stop contributions to
${\cal H}_1^{2\ell}$ and ${\cal H}_2^{2\ell}$ in the VHML, and in
section \ref{sec:corr} we further elaborate on the validity of the
approximation.  For what concerns the heaviest eigenstate $H$, general
formulae for the production cross-section can be obtained
straightforwardly with the replacements ($ -\sin\alpha \rightarrow
\cos\alpha,\,\cos\alpha \rightarrow \sin\alpha$) in eqs.~(\ref{ggh})
and (\ref{real}).  However, depending on the choice of MSSM
parameters, it might not be possible to rely on the assumption that
$H$ is much lighter than the particles running in the loops.



\section{Explicit two-loop results in the vanishing Higgs-mass limit}
\label{sec:results}

In this section we derive explicit and (relatively) compact formulae,
valid for arbitrary parameters in the stop sector, for the
contributions of two-loop diagrams involving top quarks and/or stop
squarks to the form factors for the interaction of a CP-even Higgs
boson with two gluons in the VHML. Besides providing an independent
check of the results of ref.~\cite{HS2}, our formulae can be easily
modified to allow for different renormalization schemes for the input
parameters. Furthermore we discuss how our results for the
contributions involving the stop squarks can be adapted to the squarks
of other flavors.

\subsection{Derivation of the two-loop top/stop contributions}
\label{subsec:calculation}

The starting point of our derivation is the low-energy theorem (LET)
for Higgs interactions \cite{LET1} (see also ref.~\cite{LET2}),
relating the amplitude ${\cal M}(X,\phi)$ for a generic particle
configuration $X$ plus an external Higgs boson $\phi$ of vanishing
momentum to the corresponding amplitude without the external Higgs
boson, ${\cal M}(X)$. The LET can be stated as follows: the amplitude
${\cal M}(X,\phi)$ can be obtained by considering ${\cal M}(X)$ as a
field-dependent quantity via the dependence of the relevant parameters
(masses and mixing angles) on $\phi$. The first term in the expansion
of ${\cal M}(X)$ in the Higgs field, evaluated at the minimum of the
Higgs potential, corresponds to ${\cal M}(X,\phi)$. Strictly speaking,
in case ${\cal M}(X)$ contains infrared (IR) divergent terms the
theorem applies to the IR-safe part of the two amplitudes. If $\phi$
represents a pseudoscalar Higgs boson and $X$ a pair of vector bosons,
an additional contribution to ${\cal M}(X,\phi)$ is induced by the
axial-current anomaly. This contribution cannot be expressed in terms
of derivatives of ${\cal M}(X)$ and must be computed explicitly.

To derive the CP-even Higgs-gluon-gluon form factors in the VHML we
apply the LET, identifying ${\cal M}(X)$ with the gluon self-energy in
the background-field gauge \cite{bground}. Then, the top/stop
contributions to the form factors ${\cal H}_i~(i=1,2)$ are given by
\be
\left.{\cal H}_i \,\right|^{\rm top/stop}_{m_\phi^2=0} 
~=~ \frac{2\pi v}{\alpha_s\,T_F}\;\frac{\partial \Pi^{t}(0)}{\partial S_i}~,
\label{eq:hhat}
\ee
where $v\equiv (v_1^2+v_2^2)^{\,1/2} \approx 246~{\rm GeV}$ is the electroweak
symmetry breaking parameter, $S_i~(i=1,2)$ are the CP-even parts of
the neutral components of the two MSSM Higgs doublets and
$\Pi^{t}(q^2)$ denotes the top/stop contribution to the transverse
part of the adimensional (i.e.~divided by $q^2$) self-energy of the
gluon. In analogy with the discussion in ref.~\cite{DSZ}, the
dependence of the latter on the Higgs fields $S_i$ can be identified
through the field dependence of the top and stop masses and of the
stop mixing angle. The self-energy depends also upon a fifth
field-dependent parameter related to the phase difference between the
top and stop fields. However, this parameter is relevant only when one
consider derivatives with respect to the CP-odd fields, thus it can be
ignored in our case. As in ref.~\cite{DSZ}, a lengthy but
straightforward application of the chain rule allows us to express the
functions $F_t\,,~G_t\,,\widetilde F_t$ and $\widetilde G_t$ appearing
in eqs.~(\ref{eq:H1}, \ref{eq:H2}) and in eq.~(\ref{eq:Dq}) as
combinations of the derivatives of the gluon self-energy with respect
to the top and stop masses and to the stop mixing angle. In
particular, we find for the two-loop parts of the functions
\bea
F_t^{2\ell} &=& \frac{\partial \gam}{\partial \tu}
-\frac{\partial \gam}{\partial \td}
-\frac{4\,\cdt^2}{\tu-\td}\,\frac{\partial \gam}{\partial \cdt^2}
\label{Fderiv}\,,\\
\label{Gderiv}G_t^{2\ell} &=& \frac{\partial \gam}{\partial \tu}
+\frac{\partial \gam}{\partial \td}
+\frac{\partial \gam}{\partial \t}\,,\\
\widetilde F_t^{2\ell} &=& \frac{\partial \gam}{\partial \tu}
-\frac{\partial \gam}{\partial \td}\label{Ftderiv}
+\frac{4\,\sdt^2}{\tu-\td}\,\frac{\partial \gam}{\partial \cdt^2}\,,\\
\widetilde G_t^{2\ell} &=& \frac{\partial \gam}{\partial \tu}
+\frac{\partial \gam}{\partial \td}\,,\label{Gtderiv}
\eea
where, to reduce clutter, we used the shortcut $\gam \equiv
(2/T_F)\, \Pi^{2\ell,\,t}(0)$, after decomposing the gluon self-energy in
one- and two-loop parts as
\be
\Pi(q^2) ~=~ \frac{\alpha_s}{\pi}\,\Pi^{1\ell}(q^2) 
~+~ \left(\frac{\alpha_s}{\pi}\right)^2\,\Pi^{2\ell}(q^2)
~+~{\cal O}(\alpha_s^3)~.
\ee

We computed the contributions to the gluon self-energy from the
two-loop diagrams that involve top and/or stops with the help of {\tt
FeynArts} \cite{feynarts}, using a version of the MSSM model file
adapted to the background field gauge. After isolating the transverse
part of the self-energy with a suitable projector, we Taylor-expanded
it in powers of the squared external momentum $q^2$.  The zeroth-order
term of the expansion vanishes as a consequence of gauge invariance,
while the first-order term corresponds indeed to $\Pi^{2\ell,\,t}
(0)$. We evaluated the two-loop vacuum integrals using the results of
ref.~\cite{tausk}. Finally, we computed all the
derivatives\footnote{In ref.~\cite{tausk} the two-loop vacuum
integrals are expressed in terms of a function
$\Phi(m_1^2,m_2^2,m_3^2)$, whose derivatives can be easily obtained
using the results of appendix A of ref.~\cite{DS}.} of $\gam$ that
enter eqs.~(\ref{Fderiv})--(\ref{Gtderiv}).

We performed the two-loop computation using dimensional regularization
(DREG) and modified minimal subtraction ($\msbar$). However, it is
convenient to express our results in terms of parameters renormalized
in the $\drbar$ scheme, which is based on dimensional reduction (DRED)
and preserves the supersymmetric Ward identities and relations. The
conversion of the parameters from the $\msbar$ scheme to the $\drbar$
scheme was discussed in ref.~\cite{MV}. In particular, the $\drbar$
Higgs-quark-quark Yukawa couplings differ from their $\msbar$
counterparts by a finite one-loop shift which, when inserted in the
one-loop part of a calculation, induces an additional two-loop
contribution. On the other hand, the couplings of the Higgs bosons to
squarks, as far as strong corrections are concerned, are the same in
both schemes, and they are related by supersymmetry to the
corresponding $\drbar$ Yukawa couplings. Specializing to our
calculation, only the top contribution to ${\cal H}_2^{1\ell}$ is
going to induce an additional two-loop contribution when the top
Yukawa coupling is converted from its $\msbar$ value to its $\drbar$
value, while the stop contributions to ${\cal H}_1^{1\ell}$ and ${\cal
H}_2^{1\ell}$ can be directly identified as expressed in terms of
$\drbar$ parameters. However, as can be seen from eqs.~(\ref{eq:H2},
\ref{eq:G1l}, \ref{Glimit}), in the VHML the top-quark contribution to
${\cal H}_2^{1\ell}$ goes to a constant, i.e.~it does not actually
depend on the top Yukawa coupling. Therefore we need not introduce any
additional contribution\footnote{Conversely, ref.~\cite{babis2} shows
that, if the Higgs-squark-squark coupling is expressed in terms of the
$\msbar$ Yukawa coupling, an additional two-loop contribution must be
introduced.}  to the two-loop results obtained using DREG.

In the explicit formulae for the derivatives of $\gam$ we identify the
contributions of diagrams with gluons ($g$), with strong,
D-term-induced quartic stop couplings ($4\tilde t$), and with gluinos
($\tilde g$). Assuming that the parameters in ${\cal H}_1^{1\ell}$ and
${\cal H}_2^{1\ell}$ are expressed in the $\drbar$ scheme, the
non-vanishing contributions of the two-loop diagrams with gluons read
\bea
\label{eq:glu}
\frac{\partial \gam^{g}}{\partial m_t^2}&=&~\,
\frac{1}{2\,m_t^2}\,\left(C_F - \frac{5\, C_A}{3}\right)\,,\\
&&\nn\\
\frac{\partial \gam^{g}}{\partial \ti}&=&\!\!-\frac{1}{2\,\ti}
\,\left(\frac{3\,C_F}{4} + \frac{C_A}{6}\right)\,,
\label{eq:glusqua}
\eea
where $i=1,2$. Under the same assumption for the renormalization of
the input parameters, the non-vanishing contributions of the two-loop
diagrams that involve strong quartic stop couplings read
\bea
\label{eq:qua1}
\frac{\partial \gam^{4\tilde t}}{\partial \tu}&=&-\frac{C_F}{24}\,\left[
\frac{\cdt^2\,\tu+\sdt^2\,\td}{\tuq}
+\frac{\sdt^2}{\tuq\,\td}\,\left(
\tuq\,\ln\frac{\tu}{Q^2}-\tdq\,\ln\frac{\td}{Q^2}\right)\right]~,\\
&&\nn\\
\label{eq:qua2}
\frac{\partial \gam^{4\tilde t}}{\partial \td}&=&-\frac{C_F}{24}\,\left[
\frac{\cdt^2\,\td+\sdt^2\,\tu}{\tdq}
+\frac{\sdt^2}{\tdq\,\tu}\,\left(
\tdq\,\ln\frac{\td}{Q^2}-\tuq\,\ln\frac{\tu}{Q^2}\right)\right]~,\\
&&\nn\\
\label{eq:qua3}
\frac{\partial \gam^{4\tilde t}}{\partial \cdt^2}&=&-\frac{C_F}{24}
\,\left[\frac{(\tu-\td)^2}{\tu\,\td}
-\frac{\tu-\td}{\td}\,\ln\frac{\tu}{Q^2}
-\frac{\td-\tu}{\tu}\,\ln\frac{\td}{Q^2}\right]~,
\eea
where $Q$ is the renormalization scale at which the $\drbar$
parameters in the one-loop form factors are expressed. Finally, the
contributions of the two-loop diagrams that involve gluinos are
somewhat longer and we report them in the appendix.

\subsection{Input parameters and renormalization schemes}
\label{sec:schemes}

To account for the case in which the parameters are expressed in a
renormalization scheme different from $\drbar$, we just have to shift
the parameters appearing in the one-loop part of the form factors,
after taking the limit of zero Higgs mass in the one-loop functions
${\mathcal G}^{1\ell}_{0}$ and ${\mathcal G}^{1\ell}_{1/2}$.  Since we
are focusing on the two-loop QCD corrections we need to provide a
renormalization prescription only for the top and stop masses, for the
stop mixing angle, and for the soft SUSY-breaking trilinear coupling
$A_t$. Indicating, generically, a quantity in the $\drbar$ scheme as
$x^{\smalldrbar}$, and the same quantity in a generic scheme $R$ as
$x^\smallr$, we can write the one-loop relation as $x^{\smalldrbar} =
x^\smallr + \delta x$. Then, if the one-loop form factors are
evaluated in terms of $R$ quantities, the two-loop functions in
eqs.~(\ref{Fderiv})--(\ref{Gtderiv}) must be replaced by
\bea 
F_t^{2\ell} & \longrightarrow & F_t^{2\ell} ~+~
\frac{\pi}{6\,\alpha_s}\,\left[
\frac{\delta\tu}{\tuq}-\frac{\delta\td}{\tdq}-\left(
\frac{\delta\mt}{\mt}+\frac{\delta\sdt}{\sdt}\right)
\,\left(\frac{1}{\tu}-\frac{1}{\td}\right)\right]~,\label{shiftF}\\
&&\nn\\
G_t^{2\ell} & \longrightarrow & G_t^{2\ell} ~+~
\frac{\pi}{6\,\alpha_s}\,\left[
\frac{\delta\tu}{\tuq}+\frac{\delta\td}{\tdq}-
2\,\frac{\delta\mt}{\mt}\,\left(\frac{1}{\tu}+\frac{1}{\td}\right)\right]~,\\
&&\nn\\
\label{shiftFt}
\widetilde F_t^{2\ell} & \longrightarrow & \widetilde F_t^{2\ell} ~+~
\frac{\pi}{6\,\alpha_s}\,\left[
\frac{\delta\tu}{\tuq}-\frac{\delta\td}{\tdq}
- \frac{\delta\cdt}{\cdt}\,\left(\frac{1}{\tu}-\frac{1}{\td}\right)\right]~,\\
&&\nn\\
\label{shiftGt}
\widetilde G_t^{2\ell} & \longrightarrow & \widetilde G_t^{2\ell} ~+~
\frac{\pi}{6\,\alpha_s}\,\left[
\frac{\delta\tu}{\tuq}+\frac{\delta\td}{\tdq}\right]~.
\eea
In addition, the two-loop form factor ${\cal H}_2^{2\ell}$ gets a
contribution originating from the shift in $A_t$:
\be
{\cal H}_2^{2\ell} \longrightarrow {\cal H}_2^{2\ell} ~-~
\frac{\mt\,\sdt}{s_\beta}\,
\frac{\pi}{6\,\alpha_s}
\,\left(\frac{1}{\tu}-\frac{1}{\td}\right)\,\delta A_t~.\label{shiftA}
\ee

A commonly adopted renormalization scheme for the input parameters is
the so-called on-shell (OS) scheme, in which the top and stop
parameters are related to physical quantities. In the OS scheme the
top and stop masses are defined as the poles of the corresponding
propagators, and the shifts w.r.t.~the $\drbar$ scheme are
\be
\delta \mt = {\rm Re}\,\widehat{\Sigma}_t(\mt)~,~~~~~~
\delta \tu = {\rm Re}\,\widehat{\Pi}_{11}(\tu)~,~~~~~~
\delta \td = {\rm Re}\,\widehat{\Pi}_{22}(\td)~,
\label{shiftmass}
\ee
where $\widehat{\Sigma}_t(\mt)$ and $\widehat{\Pi}_{ii}(\ti)$ denote
the finite parts of the self-energies of top and stops, respectively,
each computed at an external momentum equal to the corresponding
particle's mass. Explicit formulae for the various shifts in
eq.~(\ref{shiftmass}) can be found, e.g., in eqs.~(B.2)--(B.4) of
ref.~\cite{DSZ}. For the shift in the stop mixing angle several OS
definitions are possible. We choose $\delta \theta_t$ in such a way
that it cancels the anti-hermitian part of the stop wave-function
renormalization (w.f.r.) ~matrix, leading to \cite{mixing}
\be
\delta \theta_t = \frac12\,
\frac{\widehat{\Pi}_{12}(\tu)+\widehat{\Pi}_{12}(\td)}{\tu-\td}~,
\label{dthus}
\ee
where $\widehat{\Pi}_{12}(q^2)$ denotes the finite part of the
off-diagonal self-energy of the stops, and is given in eq.~(B.7) of
ref.~\cite{DSZ}. Finally, the trilinear coupling $A_t$ is related to
the other parameters in the top/stop sector by
\be
\sdt = \frac{2\,\mt\,(A_t+\mu\,\cot\beta)}{\tu-\td}~,
\label{sdt}
\ee
therefore, since $\mu$ and $\tan\beta$ do not get any ${\cal
O}(\alpha_s)$ correction, the shift for $A_t$ is not an independent
quantity and it can be expressed in terms of the other shifts:
\be
\delta A_t ~=~ \left(\frac{\delta \tu-\delta\td}{\tu-\td}
+\frac{\delta\sdt}{\sdt}-\frac{\delta\mt}{\mt}\right)(A_t+\mu\,\cot\beta)~.
\ee
We have verified that, in the OS scheme, the shifts in
eqs.~(\ref{shiftF})--(\ref{shiftA}) cancel the explicit dependence of
${\cal H}_1^{2\ell}$ and ${\cal H}_2^{2\ell}$ on the renormalization
scale $Q$.

We compared our results with those of the public computer code {\tt
evalcsusy.f}, which is based on the results of ref.~\cite{HS2}. The
code provides the one- and two-loop parts of the Wilson coefficient
for the Higgs-gluon-gluon operator in the effective Lagrangian, see
eq.~(2.1) and (2.5) of ref.~\cite{HS2}, using an OS renormalization
scheme for the parameters in the top/stop sector. After taking into
account the different renormalization prescription for the stop mixing
angle, the different convention for the sign of $\mu$ in
eq.~(\ref{sdt}), and an overall multiplicative factor in the
normalization of the coefficients, we find {\em perfect} numerical
agreement between our results and those of ref.~\cite{HS2}. However,
we would like to comment on the renormalization prescription for the
stop mixing angle adopted in ref.~\cite{HS2} (see also
ref.~\cite{paolosven}). Their counterterm is given by
\be
\delta \theta_t = 
\frac{{\Pi}_{12}(q_0^2)}{\tu-\td}~,
\label{dththem}
\ee
where $q_0$ is an arbitrary external momentum (a free input parameter
of {\tt evalcsusy.f}) chosen to be of the order of the stop
masses. The divergent part of the counterterm for $\theta_t$ is indeed
compelled to have the form of eq.~(\ref{dthus}) -- with the finite
part of the self-energy replaced by the divergent part -- by the
requirement that it cancel the poles of the anti-hermitian part of the
stop w.f.r.~matrix. The renormalization prescription given in
eq.~(\ref{dththem}) fulfills this requirement in the case of the QCD
corrections, because the divergent part of the ${\cal O}(\alpha_s)$
contribution to $\Pi_{12}(q^2)$ does not depend on $q^2$. In general,
however, this is not the case, unless $q_0^2 =
(\tu+\td)/2$. Therefore, we find it preferable to stick to the
``symmetrical'' prescription for $\delta \theta_t$, eq.~(\ref{dthus}),
which can be more naturally applied to other loop corrections.

\subsection{Contributions from squarks of other flavors}
\label{sec:bot}

The results presented in the previous subsections are valid in the
limit in which the Higgs boson mass is negligible with respect to the
masses of the particles circulating in the loops. Therefore, care must
be taken in extending the results derived for the top/stop
contributions to the contributions of quarks and squarks of other
flavors. In the case of the bottom/sbottom contributions the general
formulae of section \ref{sec:general} hold. However, the VHML can be
strictly applied only to the two-loop contributions arising from
diagrams with sbottoms and gluons, and to those arising from diagrams
with quartic sbottom couplings. Results valid in the $\drbar$ scheme
can be obtained for the former from eq.~(\ref{eq:glusqua}) and for the
latter from eqs.~(\ref{eq:qua1})--(\ref{eq:qua3}), with the trivial
substitution $\tilde t \rightarrow \tilde b$ in the squark masses and
mixing angle.  Obviously, the VHML cannot be applied to the
contributions arising from two-loop diagrams with bottom quarks and
gluons, but the exact results for those contributions are available in
the literature \cite{SDGZ,babis1,ABDV}. For what concerns the
contributions of two-loop diagrams with bottom, sbottom and gluino,
the VHML can only be applied under the further approximation that the
bottom mass and the left-right mixing in the sbottom sector are set to
zero (i.e.~$m_b=\theta_b=0$), effectively killing the Yukawa-induced
interactions between Higgs bosons and bottom (s)quarks. Since the
left-right sbottom mixing contains a term proportional to
$m_b\tan\beta$, this is not a good approximation when $\tan\beta$ is
large enough to offset the smallness of $m_b$.

When the bottom Yukawa coupling is neglected, the only diagrams that
give a contribution to the form factors for the Higgs-gluon-gluon
interaction are those in which the Higgs boson couples to the sbottoms
through the electroweak, D-term-induced interaction. These diagrams
contribute to the function $D_b$ that appears in eqs.~(\ref{eq:H1},
\ref{eq:H2}) and is further decomposed into two functions $\widetilde
F_b$ and $\widetilde G_b$ in eq.~(\ref{eq:Dq}). The expressions in
eqs.~(\ref{Ftderiv}) and (\ref{Gtderiv}) for the two-loop part of the
functions simplify to:
\be
\widetilde F_b^{2\ell} ~=~ \frac{\partial \gam}{\partial \bl}
-\frac{\partial \gam}{\partial \br}~,~~~~~~~~~
\widetilde G_b^{2\ell} ~=~ \frac{\partial \gam}{\partial \bl}
+\frac{\partial \gam}{\partial \br}~,
\label{dtermsb}
\ee
where, in the absence of left-right mixing, the sbottom mass
eigenstates $\sbu$ and $\sbd$ are identified with $\tilde b_L$ and
$\tilde b_R$, respectively. The $\drbar$ contributions from the
two-loop diagrams with gluons and with quartic sbottom coupling can
again be read off eqs.~(\ref{eq:glusqua})--(\ref{eq:qua2}) after
replacing $\tilde t \rightarrow \tilde b$ and setting
$\theta_b=0$. The contribution of the two-loop diagram with gluino,
sbottom and (massless) bottom reads~\footnote{{\bf Note Added:} We
  later found that eq.~(42) is incorrect. The term proportional to
  $C_F$ in the contribution of the two-loop diagram with gluino,
  sbottom and bottom can in fact be obtained by setting $\theta_b=0$
  in eq.~(40) of ref.~\cite{bottom}.}
\bea
\frac{\partial \gam^{\tilde g}}{\partial \bi} &=&
\!\frac{\g}{6\,\biq}\,C_F\,\left[
1 - \log\frac{\g}{Q^2} + \frac{\biq}{(\g-\bi)^2}\,
\left(2 + \frac{\g+\bi}{\g-\bi}\,\log\frac{\bi}{\g}\right)\,\right]\nn\\
&&\nn\\
&&-
\frac{C_A}{12\,(\g-\bi)}\,\left(1+\frac{\g}{\g-\bi}\,\log\frac{\bi}{\g}\right)~,
\label{zerolimit}
\eea
where $i = { L,R}$. If the sbottom masses appearing in the one-loop
part of the form factor are taken as the physical ones, the functions
$\widetilde F_b^{2\ell}$ and $\widetilde G_b^{2\ell}$ must be shifted
as in eqs.~(\ref{shiftFt}) and (\ref{shiftGt}), neglecting the term
proportional to $\delta c_{2\theta_b}$. It is useful to remark that,
as is clear from eq.~(\ref{eq:Dq}), there is a partial cancellation
between the sbottom contributions in $D_b$ and the corresponding stop
contributions in $D_t$. Therefore, the sbottom contributions
controlled by the electroweak gauge couplings must be taken into
account even when $\tan\beta$ is small and the bottom Yukawa coupling
can be neglected.

Finally, for the squarks of the first two generations the
approximation of neglecting the Yukawa couplings is always
satisfactory. The remaining D-term-induced contributions to the form
factors can be obtained by trivially adapting the results in
eqs.~(\ref{eq:glusqua})--(\ref{eq:qua2}) and (\ref{zerolimit}).



\section{On the validity of the vanishing Higgs-mass limit}
\label{sec:corr}

In the previous section we presented analytic results for the two-loop
form factors ${\cal H}_i^{2\ell}$ valid in the VHML.  As already
mentioned, we can expect this approximation to be quite good for the
lightest CP-even Higgs boson $h$, while for the heaviest Higgs boson
$H$ it is probably less accurate. To put this expectation on a more
solid ground we computed directly the two-loop top/stop contribution
to the Higgs-gluon-gluon amplitude via a Taylor expansion in the
external Higgs momentum up to terms of ${\cal O}(m_\phi^2/M^2)$ --
where $\phi = h,H$, and $M$ denotes generically the masses of the
heavy particles in the loop (i.e.~top, stops and gluino). The validity
of the Taylor expansion is restricted to Higgs masses below the first
threshold that is encountered in the diagrams. This is always the case
for $h$, while for $H$ this situation is realized only in specific
regions of the parameter space.

We computed the Higgs-gluon-gluon amplitude following the same
strategy employed for the calculation of the gluon self-energy, see
section \ref{subsec:calculation}. The zeroth-order term in the Taylor
expansion reproduces the result that we obtained via the LET, while
the ${\cal O}(m_\phi^2/M^2)$ term in the expansion gives the first
correction to the VHML. For simplicity we neglected all the (small)
D-term-induced electroweak contributions. We expressed our results in
the OS renormalization scheme outlined in section
\ref{sec:schemes}. We remark that, when converting to the OS scheme
the results obtained originally in the $\msbar$ scheme, we must
introduce additional two-loop contributions of ${\cal
O}(m_\phi^2/M^2)$, originating from the shifts in the parameters that
appear in the ${\cal O}(m_\phi^2/M^2)$ parts of the one-loop form
factors -- see eq.~(\ref{Glimit}). We checked that the additional
contributions cancel the explicit renormalization-scale dependence of
the ${\cal O}(m_\phi^2/M^2)$ part of the two-loop form factors. The
analytic expressions for the ${\cal O}(m_\phi^2/M^2)$ corrections are
very long and we do not report them.

\begin{figure}[p]
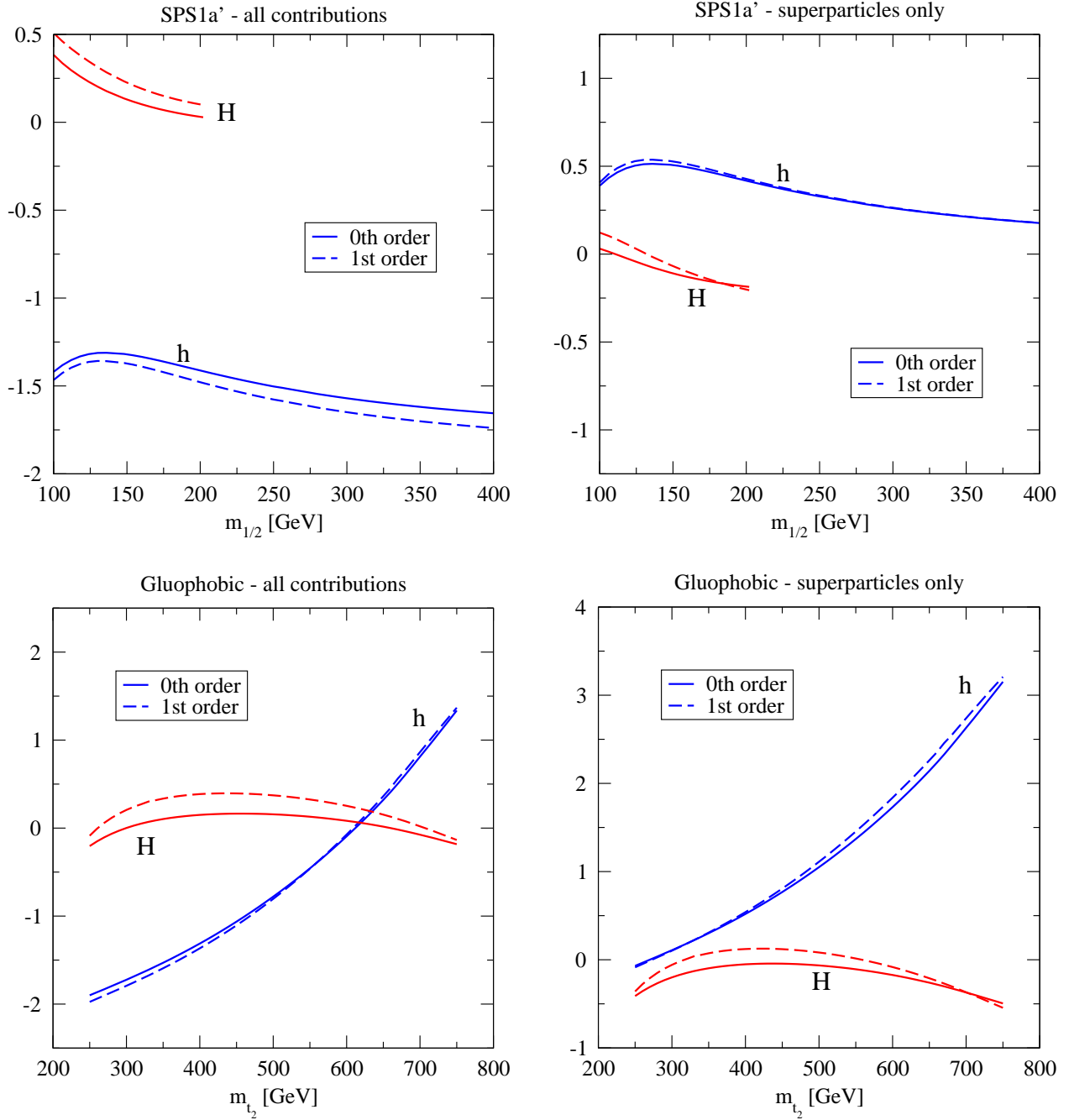

\begin{center}
\mbox{
\epsfig{figure=sps1ap_all_OS.eps,width=8cm}~~~~
\epsfig{figure=sps1ap_new_OS.eps,width=8cm}}

\vspace{5mm}

\mbox{
~~\epsfig{figure=glphob_all_OS.eps,width=7.75cm}~~~~~~
\epsfig{figure=glphob_new_OS.eps,width=7.75cm}}
\end{center}
\vspace{-2mm}
\caption{\sf Two-loop form factors ${\cal H}^{2\ell}_{h}$ and ${\cal
H}^{2\ell}_{H}$ in the SPS1a$^\prime$ scenario (upper plots) and in
the gluophobic scenario (lower plots). The plots on the right include
only the superparticle contributions. In each plot, the solid lines
refer to the result obtained in the limit of vanishing Higgs mass,
while the dashed lines include also terms of ${\cal
O}(m_\phi^2/M^2)$. The lines for ${\cal H}^{2\ell}_{H}$ in the
SPS1a$^\prime$ scenario are truncated where $m_H \approx 2\,m_t$.}
\label{plot}
\end{figure}

Similarly to ref.~\cite{HS2}, we present our results in two representative
MSSM scenarios. The first scenario is the so-called SPS1a$^\prime$
slope \cite{sps1ap}, in which the soft SUSY-breaking parameters at the
GUT scale are related as
\be
m_0 ~=~ 0.28~m_{1/2}~,~~~~~~A_0~=~-1.2~m_{1/2}~,
\label{sps1ap}
\ee
where $m_0$ and $m_{1/2}$ are universal SUSY-breaking masses for
scalars and gauginos, respectively, $A_0$ is a universal
Higgs-sfermion-sfermion interaction term, and the other relevant
parameters are $\tan\beta = 10$ and $\mu<0$ (with our sign
convention). We vary the GUT-scale gaugino mass $m_{1/2}$ between 100
GeV and 400 GeV, and the other SUSY-breaking parameters as in
eq.~(\ref{sps1ap}) above.  To compute the superparticle masses, we
evolve the soft SUSY-breaking parameters down to the weak scale using
the public computer code {\tt SoftSusy} \cite{softsusy}. In the second
scenario the light Higgs boson is ``gluophobic'', i.e.~the top quark
contribution to the Higgs-gluon coupling is largely canceled by the
contribution of a light stop. The scenario is defined directly in
terms of weak-scale parameters, which we choose as
\be 
\tan\beta = 10,~~~\theta_t = \frac{\pi}{4},~~~m_{A}= 300\,{\rm GeV},
~~~\mu = -500\,{\rm GeV},~~~\mg = 500\,{\rm GeV},~~~m_{\stu}= 200\,{\rm GeV},
\ee
while $m_{\std}$ is varied between 250 GeV and 750 GeV. In the first
scenario the CP-even Higgs masses and mixing angle are computed
directly by {\tt SoftSusy}, while in the second scenario we compute
them using the two-loop ${\cal O}(\alpha_t\alpha_s)$ results of
ref.~\cite{DSZ}. In both scenarios we take $m_t = 172.6$ GeV
\cite{topmass}.

The four plots in figure \ref{plot} show our results for the two-loop
form factors ${\cal H}^{2\ell}_{h}$ and ${\cal H}^{2\ell}_{H}$, which
we define as
\be
{\cal H}^{2\ell}_{h} ~=~ T_F\,\left(-\sin\alpha \,{\mathcal H}^{2\ell}_1 +
\cos\alpha \,{\mathcal H}^{2\ell}_2 \right)~,~~~~~~
{\cal H}^{2\ell}_{H} ~=~ T_F\,\left(\cos\alpha \,{\mathcal H}^{2\ell}_1 +
\sin\alpha \,{\mathcal H}^{2\ell}_2 \right)~.
\ee
The two upper plots refer to the SPS1a$^\prime$ scenario, while the
two lower plots refer to the gluophobic scenario. For each scenario,
the plot on the left shows the complete two-loop top/stop
contributions to the form factors, while the plot on the right shows
only the contributions of the diagrams that include
superparticles. Finally, for each Higgs boson $\phi = h,H$\,, the
solid line corresponds to the result obtained for ${\cal
H}^{2\ell}_{\phi}$ in the VHML, while the
dashed line includes also the contribution of the first-order term in
the expansion in powers of $m_\phi^2/M^2$.

It can be seen from figure \ref{plot} that, for the lightest Higgs
boson $h$, the corrections of ${\cal O}(m_h^2/M^2)$ to the results
obtained for ${\cal H}^{2\ell}_{h}$ in the VHML are quite small, which
should not come as a surprise since $m_h$ is always considerably
smaller than $m_t$. In the SPS1a$^\prime$ scenario, which as $m_{1/2}$
increases is characterized by relatively heavy superparticles, the
comparison between the plots on the left and right sides shows that
the bulk of the corrections is contained in the diagrams with top
quarks and gluons, for which a complete analytic result (i.e., valid
for any value of the top and Higgs masses) is available
\cite{SDGZ,babis1,ABDV}. In the gluophobic scenario there is still a
small Higgs-mass dependence in the superparticle contribution, due to
the presence of a relatively light stop. In summary, it appears to be
quite safe to approximate the two-loop top and stop contributions to
${\cal H}^{2\ell}_{h}$ with the results obtained via the LET in
section \ref{sec:results}. In case a more refined approximation is
required, one can implement the complete result for the top/gluon
contribution by replacing
\be
\frac{\partial \gam^{g}}{\partial m_t^2} ~\longrightarrow~
\frac{1}{2\,m_t^2}\,\biggr[C_F\, {\cal G}^{(2\ell,C_R)}_{1/2} + C_A\, 
{\cal G}^{(2\ell,C_A)}_{1/2}\,\biggr]~,
\ee
in the gluonic part of the function $G_t^{2\ell}$ defined in
eq.~(\ref{Gderiv}).  The functions ${\cal G}^{(2\ell,C_R)}_{1/2}$ and
${\cal G}^{(2\ell,C_A)}_{1/2}$ in the OS renormalization scheme are
defined in eqs.~(2.15) and (3.8) of ref.~\cite{ABDV}, respectively.

The situation is quite different for the heaviest Higgs boson $H$. As
we mentioned before, the limit of vanishing $m_H$ can only be
considered if $m_H$ is smaller than the lowest threshold appearing in
the loops. Indeed, the curves for ${\cal H}^{2\ell}_{H}$ in the
SPS1a$^\prime$ scenario are truncated around $m_{1/2} = 200$ GeV,
where $m_H$ approaches $2\,m_t$. On the other hand, it can be seen
from figure \ref{plot} that, in the regions of the parameter space
where the limit of vanishing $m_H$ can be applied at all, the
resulting approximation is not exceedingly bad even for $m_H$ as large
as 300 GeV. It should however be kept in mind that, in the case of
large $\tan \beta$, the VHML approximation cannot be applied even if
$m_H$ is relatively small, because the $H$ boson has enhanced
couplings to the bottom (s)quarks (see the discussion in section
\ref{sec:bot}). In that case a full computation is unavoidable
\cite{babis2}.



\section{Gluonic and photonic Higgs decays}
\label{sec:decays}

The results of the previous sections can be directly applied to the
NLO computation of the gluonic decay widths of the CP-even Higgs
bosons in the MSSM.  We specialize to the decay width of the $h$
boson, but the formulae presented below can be applied also the decay
of $H$ using the replacements indicated at the end of section
\ref{sec:general} (with the same restrictions outlined there and at
the end of the previous section).

At NLO in QCD the decay width of the lightest Higgs boson in two
gluons reads
\be
\label{hgluglu}
\Gamma(h\rightarrow gg) ~=~
\frac{G_\mu\,\alpha_{s}(\muR)^2\,m_h^3}{16\,\sq2\,\pi^3}~
\biggr|T_F \left( -\sin\alpha\,{\cal H}^{1\ell}_1 
+ \cos\alpha\,{\cal H}^{1\ell}_2 \right) \biggr|^2 
\left(1+\frac{\alpha_s}{\pi}\,{\cal C}\,\right)~,
\ee
where ${\cal C} = {\cal C}_{ virt} +{\cal C}_{ggg} + {\cal C}_{g q
\bar{q}}$ includes contributions from the two-loop virtual corrections
and from the one-loop real radiation processes $h \to ggg,\: h \to g q
\bar{q}$.  The contribution of the two-loop virtual corrections is
straightforwardly obtained from the results of the previous sections:
\be
\label{Cvirt}
{\cal C}_{virt} ~=~ 
C_A \, \frac{~\pi^2}3 + \beta_0\, \ln \left( \frac{\muR^2}{m_h^2} \right) 
+\left(\frac{-\sin\alpha \,{\mathcal H}^{2\ell}_1 
+\cos\alpha \,{\mathcal H}^{2\ell}_2}{
-\sin\alpha \,{\mathcal H}^{1\ell}_1 
+ \cos\alpha \,{\mathcal H}^{1\ell}_2 } + {\rm h.c.} \right)~,
\ee
while the contributions of the real radiation processes are, in the
VHML,
\be
{\cal C}_{ggg} ~=~ 
- C_A \left(\frac{~\pi^2}3 - \frac{73}{12} \right)~,~~~~~~~~ 
{\cal C}_{g q \bar{q}} ~=~ -\frac76 N_f~,
\ee
where $N_f$ is the number of light quark species, with the quarks
treated as massless particles. 

Considering for illustration the two scenarios described in section
\ref{sec:corr}, we found that the numerical value of the last term
(within parentheses) in the r.h.s.~of eq.~(\ref{Cvirt}) is
approximately $-2.6\, C_F + 2.5 \,C_A$ in the SPS1a$^\prime$ scenario
with $m_{1/2}$ = 200 GeV, and $-4.5 \,C_F + 2.8 \,C_A $ in the
gluophobic scenario with $m_{\tilde t_2}$ = 500 GeV. For the heaviest
Higgs boson the corresponding values are $-2.9 \,C_F + 1.4 \,C_A$
(SPS1a$^\prime$) and $-1.1 \,C_F + 0.9 \,C_A$ (gluophobic). In both
scenarios the superparticle contribution to the coefficient of $C_F$
is comparable to the corresponding SM contribution, while the
superparticle contribution to the coefficient of $C_A$ is much smaller
than its SM counterpart.

As a byproduct of our calculation, we can also provide the explicit
results for the two-loop QCD corrections to the quark/squark
contributions to the photonic Higgs decay. The partial width for the
decay of the lightest CP-even Higgs boson $h$ in two photons can be
written as
\be
\label{hgaga}
\Gamma(h\rightarrow \gamma\gamma) ~=~
\frac{G_\mu\,\alpha_{\rm em}^2\,m_h^3}{128\,\sq2\,\pi^3}~
\biggr|-\sin\alpha\,{\cal P}_1 + \cos\alpha\,{\cal P}_2\biggr|^2~,
\ee
where $\alpha_{\rm em}$ is the electromagnetic coupling and ${\cal P}_i$
($i=1,2$) are defined, in analogy to the ${\cal H}_i$ in
eq.~(\ref{ggh}), as the form factors for the coupling of the neutral,
CP-even component of the Higgs doublet $H_i$ with two photons. At one
loop, the form factors ${\cal P}_1$ and ${\cal P}_2$ receive
contributions from all the electrically charged states of the MSSM
(see, e.g., the first paper in ref.~\cite{SDGZ} for the explicit
results) and they are in general dominated by the contribution of the
diagram involving the $W$ boson. However, only the contributions
involving quarks and squarks receive QCD corrections at two loops. We
separate the one-loop part of the form factors and the two-loop QCD
corrections as
\be
{\mathcal P}_i ~=~ {\mathcal P}_i^{1\ell}
~+~ \frac{\alpha_s}{\pi} \,  {\mathcal P}_i^{2\ell}
~+~\ldots~,
\label{Pdec}
\ee
where the ellipses stand for three-loop terms of ${\cal O}(\alpha_s^2)$
and for two-loop terms controlled by other coupling constants. 

Focusing on the contributions of the third-generation quarks and
squarks, ${\cal P}_1$ and ${\cal P}_2$ can be decomposed exactly as in
eqs.~(\ref{eq:H1}) and (\ref{eq:H2}), with the following substitutions
in the r.h.s.~of eqs.~(\ref{eq:H1}) and (\ref{eq:H2})
\be
F_q \to \widehat{F}_q,~~~~G_q \to \widehat{G}_q,~~~~D_q \to\widehat{D}_q,~~~~
\lambda_t \to \frac{Q_t^2\,N_c }{\sin\beta},~~~\lambda_b \to
\frac{Q_b^2\,N_c }{\cos\beta} 
\ee
where $Q_q$ is the electric charge of the quark.  With our overall
normalization we have for the functions entering the one-loop parts of
the form factors, ${\cal P}^{1\ell}_i$,
\be
\widehat{F}_q^{1\ell}= F_q^{1\ell},~~~~\widehat{G}_q^{1\ell}  
= G_q^{1\ell},~~~~\widehat{D}_q^{1\ell} = D_q^{1\ell}~,
\ee
while for the ones entering ${\cal P}^{2\ell}_i$
\be
\widehat{F}_q^{2\ell}= \left.F_q^{2\ell}\,\right|_{C_A=0},~~~~
\widehat{G}_q^{2\ell}  = \left.G_q^{2\ell}\,\right|_{C_A=0},~~~~
\widehat{D}_q^{2\ell} = \left.D_q^{2\ell}\,\right|_{C_A=0}~,
\ee
i.e. the functions entering the two-loop parts of the form factors can
be obtained by setting $C_A=0$ in the results presented in section
\ref{sec:results} and in the appendix.

\section{Discussion}
\label{sec:concl}

The LET allowed us to derive explicit and compact analytical formulae,
for the top/stop contributions to the form factors for the interaction
of a CP-even Higgs boson with two gluons, valid in the limit in which
the mass of the Higgs boson is neglected w.r.t.~the masses of the
particles running in the loops.  By direct inspection of the first
correction to the results obtained in the VHML, we have argued that,
for the lightest MSSM Higgs boson, the VHML results provide a quite
good approximation to the full result, whereas for the heaviest Higgs
boson the approximation is less good, and can be applied only in
specific regions of the parameter space. For what concerns the sbottom
contributions to the form factors, the validity of our results is
limited to the case of small or moderate $\tan\beta$.

As mentioned in section \ref{subsec:calculation}, the form factor for
the interaction of a CP-odd Higgs boson $A$ with two gluons receives
an additional contribution from the axial-current anomaly and cannot
be computed in the same way as the form factors for the CP-even bosons
(indeed, the first derivatives of the gluon self-energy w.r.t.~the
CP-odd parts of the Higgs fields vanish at the minimum of the Higgs
potential). In the VHML the $A$-gluon-gluon form factor does not
receive any contribution beyond one loop from diagrams involving
gluons \cite{AQCD}, as a consequence of the Adler-Bardeen theorem
\cite{ABtheo}. However, there is a non-vanishing two-loop contribution
from the diagrams involving gluinos which requires an explicit
diagrammatic calculation~\cite{HH}.

Finally, the LET can also be applied to multiple Higgs boson
production \cite{LET2}. By taking multiple derivatives of the gluon
self-energy w.r.t.~the CP-even or CP-odd Higgs fields, it is indeed
possible to compute the SUSY-QCD corrections to the production of any
number of CP-even Higgs bosons and of any even number of CP-odd Higgs
bosons in the VHML. The simplest case would be the calculation of the
SUSY-QCD corrections to the pair production of both CP-even and CP-odd
Higgs bosons. An analysis of these processes, restricted to the
top-quark contributions, has been presented in ref.~\cite{pair}.

\section*{Acknowledgments}

We thank C.~Anastasiou and G.~Ridolfi for useful discussions.  This
work was supported in part by an EU Marie-Curie Research Training
Network under contract MRTN-CT-2006-035505 and by ANR under contract
BLAN07-2\_194882.

\newpage

\section*{Appendix}
\begin{appendletterA}

We provide in this appendix the explicit expressions for the
contributions of the two-loop diagrams with top, stop and gluino to
the derivatives of $\gam$. For every derivative we separate the
coefficients of the color factors $C_F$ and $C_A$ as
\be
\frac{\partial \gam^{\gl}}{\partial x_i} ~=~ 
C_F\,\frac{\partial \gam^{\gl}_{C_F}}{\partial x_i}+
C_A\,\frac{\partial \gam^{\gl}_{C_A}}{\partial x_i}~,
\ee
for $x_i\,=\,(\t\,,~\tu\,,~\td\,,~\cdt^2)$.  In
terms of the two-loop function $\Phi(x,y,z)$ defined, e.g., in the
appendix A of ref.~\cite{DS}, and using the shortcut $\Delta\,\equiv\,
\gq+\tuq+\tq - 2\, (\g\,\tu + \g \,\t + \tu\,\t)\,$, the contributions
to the various derivatives\footnote{~$\partial
\gam^{\gl}/\partial\td$ can be obtained from $\partial
\gam^{\gl}/\partial\tu$ through the replacements
$\tilde{t}_1\,\rightarrow\,\tilde{t}_2$ and $\sdt\,\rightarrow-\sdt$.} read

%
%
\bea
%
%
\frac{\partial \gam^{\gl}_{C_F}}{\partial \tu} &=&
\frac{1}{6\,\tuq\,\del^2}\,\left[
(\g+\t)\,\del^2+2\,\g\,\tuq\,(\del+10\,\g\,\t)\right]\nn\\
&&\nn\\
&-&\frac{\mg\,\sdt}{6\,\mt\,\tuq\,\del^2}\,\left[
\t\del^2+2\,\tuq\,(\del+5\,\g\,\t)\,(\g+\t-\tu)\right]\nn\\
&&\nn\\
%
%
&-&\frac{\gq}{6\,\tuq\,\del^3}\,\log\frac{\g}{\t}
~\biggr\{(\g-\t-4\,\tu)\,\del^2\nn\\
&&~~+2\,\tu\,
\left(-18\,\t\,\tu\,\del+
((3\,\t-\g)\,\del-30\,\g\,\t\,\tu)(\t-\g+\tu)\right)\nn\\
&&\nn\\
&&~~-\frac{\sdt}{\mg\,\mt}\,\left[
(2\,\tu\,(\tu+\t)+\t\,(\g-\t))\,\del^2\right.\nn\\
&&~~-2\,\t\,\tu\,\left.
\left(-9\,\g\,\tu\,\del
+((2\,\g-9\,\tu)\,\del-30\,\g\,\t\,\tu)(\g-\t+\tu)\right)
\right]\biggr\}\nn\\
&&\nn\\
%
%
&+&\frac{\g}{6\,\del^3}\,\log\frac{\tu}{\t}~\biggr\{
\del^2 + 12\,\g\,\t\,\del+(2\,\tu\,\del+60\,\g\,\t\,\tu)(\t+\g-\tu)\nn\\
&&\nn\\
&&~~-\frac{2\,\sdt}{\mg\,\mt}\,\left[
(\g+\t)\,\del^2 + \g\,\t\,(3\,\g+3\,\t+20\,\tu)\,\del+60\,\gq\,\tq\,\tu
\right]\biggr\}\nn\\
&&\nn\\
%
%
&-&\frac{1}{6\,\tuq}\,\left(\g+\t-\sdt\,\mg\,\mt\right)
\,\log\frac{\t}{Q^2}\nn\\
&&\nn\\
%
%
&+&\frac{\gq\,\t}{\tu\,\del^3}\,\Phi\,(\g\,,\,\t\,,\,\tu)~\biggr\{
(\g+\t+3\,\tu)\,\del+20\,\g\,\t\,\tu \nn\\
&&\nn\\
&&~~-\frac{\sdt}{\mg\,\mt}\,\left[
\del^2 + 2\,\g\,\t\,\del + (3\,\tu\,\del +10\,\g\,\t\,\tu)\,(\g+\t-\tu)
\right]\biggr\}~,\\
&&\nn
\eea

%
\bea
%
%
\frac{\partial \gam^{\gl}_{C_A}}{\partial \tu} &=&
\frac{1}{12\,\del^2}\,\left[
2\,\t\,\del-(\del+20\,\g\,\t)\,(\g-\t-\tu)\right]\nn\\
&&\nn\\
&+&\frac{\mt\,\sdt}{3\,\mg\,\del^2}\,\left[
2\,\g\,\del-(\del+5\,\g\,\t)\,(\t-\g-\tu)\right]\nn\\
&&\nn\\
%
%
&+&\frac{\g}{12\,\del^3}\,\log\frac{\g}{\tu}
~\biggr\{\del^2 +2\,\t\,
\left((13\,\g+9\,\t+15\,\tu)\,\del+120\,\g\,\t\,\tu\right)\nn\\
&&\nn\\
&&~~-\frac{2\,\sdt\,\mt}{\mg}\,\left[ 18\,\t\,(\t-\tu)\,\del \right.\nn\\
&&~~~~~~~~~~~~~~~~~~\left.
+\left((11\,\g-9\,\t+9\,\tu)\,\del+60\,\g\,\t\,\tu\right)\,(\g+\t-\tu)
\right]\biggr\}\nn\\
&&\nn\\
%
%
&-&\frac{\t}{12\,\del^3}\,\log\frac{\t}{\tu}~\biggr\{
12\,\t\,(\tu-\t)\,\del\nn\\
&&~~~~~~~~~~~~~~~~~~+\left((15\,\g+13\,\t+3\,\tu)\,\del
+120\,\g\,\t\,\tu\right)(\g+\t-\tu)\nn\\
&&\nn\\
&&~~-\frac{4\,\sdt}{\mg\,\mt}\,\left[
(3\,\g+\t)\,\del^2 + \g\,\t\,\left(
(9\,\g+2\,\t+21\,\tu)\,\del+60\,\g\,\t\,\tu\right)\right]\biggr\}\nn\\
&&\nn\\
%
%
&+&\frac{\g\,\t}{2\,\tu\,\del^3}\,\Phi\,(\g\,,\,\t\,,\,\tu)~\biggr\{
2\,\t\,\tu\,\del\nn\\
&&~~~~~~~~~~~~~~~~~~~~
+\left((\g+\t+2\,\tu)\,\del+20\,\g\,\t\,\tu\right)
\,(\t-\g+\tu) 
\nn\\
&&\nn\\
&&~~+\frac{\sdt}{\mg\,\mt}\,\left[
(\g-\t+\tu)\,\del^2 + 2\,\t\,\left(2\,\tu\,(\t-\tu)\,\del \right.\right.\nn\\
&&\left.\left.~~~~~~~~~~~~~~~~~~~~+ 
((\g+5\,\tu)\,\del +10\,\g\,\t\,\tu)\,(\g-\t+\tu)\right)\right]\biggr\}~,
\eea
%
%
\bea
&&\nn\\
%
%
\frac{\partial \gam^{\gl}_{C_F}}{\partial \t} &=&
-\frac{\g}{6\,\tu\,\del^2}\,\left[
4\,\tu\,\del+(\del-10\,\g\,\tu)\,(\g-\t-\tu)\right]\nn\\
&&\nn\\
&-&\!\!\!\frac{\mg\,\sdt}{12\,\mt\,\tu\,\del^2}\left[
\del^2+2\,\left(5\,\g\,\tu\,\del+((2\,\tu-\g)\,\del+10\,\g\,\t\,\tu)
\,(\g-\t+\tu)\right)\right]\nn\\
&&\nn\\
%
%
&+&\frac{\gq}{6\,\tu\,\del^3}\,\log\frac{\g}{\t}
~\biggr\{\del^2-2\,\tu\,
\left((2\,\g+3\,\t+9\,\tu)\,\del+60\,\g\,\t\,\tu\right)\nn\\
&&\nn\\
&&~~-\frac{\sdt}{2\,\mg\,\mt^3}\,\left[
(\tq-2\,\tuq+\g\,(\t+2\,\tu))\,\del^2 + \t\,\tu\,
\left(20\,\g\,\t\,\del\right.\right.\nn\\
&&~~~~~~~~~~~~~~~\left.\left.
-((25\,\tu+9\,\g+11\,\t)\,\del+120\,\g\,\t\,\tu)(\g+\t-\tu)\right)
\right]\biggr\}\nn\\
&&\nn\\
%
%
&+&\frac{\g\,\tu}{3\,\del^3}\,\log\frac{\tu}{\t}~\biggr\{
9\,\g\,\del+(\del+30\,\g\,\t)\,(\g-\t+\tu)
\nn\\
&&\nn\\
&&~~+\frac{\sdt}{2\,\mg\,\mt^3}\,\left[
(\g+3\,\t-\tu)\,\del^2 \right.\nn\\
&&~~~~~~~~~~~~~~~~~~~\left.+ \g\,\t\,\left(12\,\t\,\del+
(17\,\del+60\,\g\,\t)\,(\t-\g+\tu)\right)\right]\biggr\}\nn\\
&&\nn\\
%
%
&+&\frac{1}{6\,\tu}\,\left(1-\frac{\sdt\,\mg}{2\,\mt}
\right)\,\log\frac{\t}{Q^2}\nn\\
&&\nn\\
%
%
&-&\frac{\gq}{\del^3}\,\Phi\,(\g\,,\,\t\,,\,\tu)~\biggr\{
2\,\t\,\del+(\del+10\,\g\,\t)\,(\t-\g+\tu)
\nn\\
&&\nn\\
&&~~+\frac{\sdt}{2\,\mg\,\mt}\,\left[
\del^2 + 2\,\t\,\left(3\,\g\,\del + (3\,\del +10\,\g\,\t)\,(\g-\t+\tu)\right)
\right]\biggr\}\nn\\
&&\nn\\
&+&~~\left(\tilde{t}_1~\rightarrow~\tilde{t}_2~,~~~
\sdt~\rightarrow-\sdt\right)~,
\eea
%
%
\bea
%
%
\frac{\partial \gam^{\gl}_{C_A}}{\partial \t} &=&
-\frac{1}{12\,\del^2}\,\left[
(5\,\g+5\,\tu-\t)\,\del+40\,\g\,\t\,\tu \right]\nn\\
&&\nn\\
&-&\frac{\sdt}{12\,\mg\,\mt\,\del^2}\,\left[
10\,\t\,(\tu-\t)\,\del\right.\nn\\
&&~~~~~~~~~~~~~~~~~~\left.+
\left((8\,\t-5\,\g-2\,\tu)\,\del-20\,\g\,\t\,\tu\right)
\,(\t+\g-\tu)\right]\nn\\
%
%
&-&\frac{\g}{12\,\del^3}\,\log\frac{\g}{\t}
~\biggr\{\del^2 +12\,\t\,\tu\,\del \nn\\
&&~~~~~~~~~~~~~~~~~~+
\left((4\,\g+18\,\tu)\,\del+120\,\g\,\t\,\tu)\,(\t-\g+\tu)\right)\nn\\
&&\nn\\
&&~~+\frac{\sdt}{\mg\,\mt}\,\left[ (\g+4\,\t\,+12\,\tu)\,\del^2 
+4\,\t\,\left(16\,\tu\,(\t-\tu)\,\del \right.\right.\nn\\
&&~~~~~~~~~~~~~~~~~~\left.\left.
+((\t+24\,\tu)\,\del+30\,\g\,\t\,\tu)\,(\g-\t+\tu)\right)
\right]\biggr\}\nn\\
&&\nn\\
%
%
&-&\frac{\tu}{12\,\del^3}\,\log\frac{\tu}{\t}~\biggr\{
8\,\t\,(\tu-\t)\,\del \nn\\
&&~~~~~~~~~~~~~~~~~~~~
+ \left((19\,\g+9\,\t+3\,\tu)\,\del+120\,\g\,\t\,\tu\right)
(\g+\t-\tu)
\nn\\
&&\nn\\
&&~~-\frac{\sdt}{\mg\,\mt}\,\left[
(11\,\g+2\,\t+2\,\tu)\,\del^2 \right.\nn\\
&&\left.~~~~~~~~~~~~~~~~~~~~+ 2\,\g\,\t\,\left(
(21\,\g+3\,\t+43\,\tu)\,\del+120\,\g\,\t\,\tu\right)\right]\biggr\}\nn\\
&&\nn\\
%
%
&-&\frac{\g}{2\,\del^3}\,\Phi\,(\g\,,\,\t\,,\,\tu)~\biggr\{
\del^2+\t\,\left((7\g+7\,\tu+\t)\,\del+40\,\g\,\t\,\tu\right)
\nn\\
&&\nn\\
&&~~-\frac{\sdt}{2\,\mg\,\mt}\,\left[
(\g+3\,\t-\tu)\,\del^2 + 2\,\t\,\left(2\,\t\,(\tu-\t)\,\del\right.\right.\nn\\
&&\left.\left.~~~~~~~~~~~~~~~~~~~~+ 
((3\,\g+2\,\t+6\,\tu)\,\del +20\,\g\,\t\,\tu)
\,(\g+\t-\tu)\right)\right]\biggr\}\nn\\
&&\nn\\
&+&~~\left(\tilde{t}_1~\rightarrow~\tilde{t}_2~,~~~
\sdt~\rightarrow-\sdt\right)~,
\eea

%
\bea
\frac{\partial \gam^{\gl}_{C_F}}{\partial \cdt^2} &=&
\frac{\mt\,\mg}{12\,\sdt\,\tu\,\del}\,\left(2\,\g\,\tu-\del\right)\nn\\
&&\nn\\
&+& \frac{\mg^3}{12\,\sdt\,\mt\,\tu\,\del^2}\,
\left[4\,\t\,\tu\,\del\right.\nn\\
&&\left.~~~~~~~~~~~~~~~~~~~~~~~
+\left((2\,\tu-\t)\,\del+6\,\g\,\t\,\tu\right)\,(\t-\g+\tu)\right]
\,\log\frac{\g}{\t}\nn\\
&&\nn\\
&+&\frac{\mg\,\tu}{6\,\sdt\,\mt\,\del^2} \,
(\del+3\,\g\,\t)\,(\g+\t-\tu)\,\log\frac{\tu}{\t}
~+~\frac{\mg\,\mt}{12\,\sdt\,\tu}\,\log\frac{\t}{Q^2}\nn\\
&&\nn\\
&+&\frac{\mg^3\,\mt}{2\,\sdt\,\del^2}~
(\del+2\,\g\,\t)~\Phi\,(\g\,,\,\t\,,\,\tu)\nn\\
&&\nn\\
&+&~~\left(\tilde{t}_1~\rightarrow~\tilde{t}_2~,~~~
\sdt~\rightarrow-\sdt\right)~,\nn\\
&&\nn\\
%
%
\frac{\partial \gam^{\gl}_{C_A}}{\partial \cdt^2} &=&
\frac{\mt\,\mg}{12\,\sdt\,\del}\,(\t+\tu-\g)\nn\\
&&\nn\\
&+& \frac{\mg\,\mt}{12\,\sdt\,\del^2}\,
\left((\g+6\,\tu)\,\del+12\,\g\,\t\,\tu\right)\,\log\frac{\g}{\t}\nn\\
&&\nn\\
&-&\frac{\mt\,\tu}{12\,\sdt\,\mg\,\del^2} \,\left[
3\,\g\,\del
+\left(2\,\del+6\,\g\,\t\right)(\g-\t+\tu)\right]\,\log\frac{\tu}{\t}\nn\\
&&\nn\\
&-&\frac{\mg\,\mt}{4\,\sdt\,\del^2}\left[
(\del+2\,\g\,\t)\,(\g-\t-\tu)\right]\,\Phi\,(\g\,,\,\t\,,\,\tu)\nn\\
&&\nn\\
&+&~~\left(\tilde{t}_1~\rightarrow~\tilde{t}_2~,~~~
\sdt~\rightarrow-\sdt\right)~.
\eea

\end{appendletterA}

\end{document}